# Observation of 511 keV peak high count rate in studying (n,x) and (g,x) reactions on terbium

I. Kadenko* and N. Dzysiuk

International Nuclear Safety Center of Ukraine

(Str. Vasylkivska, 98-A, 03127, Kyiv, Ukraine)

## Abstract

Experimental investigation of (*n, x*) and (*g, x*) reactions on Tb-159 with activation technique was carried out. Tb specimens of natural composition were irradiated with (*d-d*) and (*d-t*) neutrons using NG-300 neutron generator. Additionally the series of experiments were performed with application of M-30 microtrone as a source of electrons for bremsstrahlung spectra production with end point energies 7.5, 9.5, 11, 11.5, 12, 12.5, 16.5, and 18.5 MeV. Instrumental spectra of Tb specimens were measured with HPGe and Ge(Li) spectrometers. Within the main scope of nuclear reactions research and accurate γ-spectrometry of Tb specimens a high count rate in 511 keV γ-line peak was observed. The first-priority analysis of Tb specimen impurities was done with further attempts to explain a result of observations with reference to the specific nuclear properties of Tb which could appear due to complex GDR structure. The energy threshold of the process detected was determined around 12.2 MeV. The lower estimate of cross section value for this process was assumed and calculated.

PACS number: 25.40.Fq, 24.10.-i, 25.60Dz

Key words: neutron, cross section, terbium, bremsstrahlung

## 1. Introduction

Present investigations were performed primary in accordance with present-day requests of nuclear data as the crucial components for developing nuclear reaction theory [1]. Attention was targeted on terbium as an element of the rare-earth group. Indeed, neutron-induced cross section data for Tb are of great importance both for nuclear technology development and for basic research. Another reason for selecting Tb is its special nuclear properties such as strong deformation. The number of rare-earth applications is increasing simultaneously with their nuclear corresponding data requests. Contemporary fusion-reactor design studies call for facility operation at a very low production rate of long-lived radionuclides so as to minimize the problems associated with reactor maintenance, low-level waste disposal, and reactor decommissioning [2]. From this view point, it is very important to know the cross section of the formation of long-lived nuclei produced via the (*n, x*) reaction. Due to the fact that terbium can be an impurity in potential blanket materials, it must be examined carefully [1, 3]. The neutron activation technique is one of the most applicable for such measurements. In order to get comprehensive knowledge about terbium's behavior, separate measurements were made with both incident neutrons and bremsstrahlung. It also should be emphasized that the study of the GDR in statically deformed nuclei of Tb provides both a sensitive test of the classical hydrodynamic model of the nucleus and means of obtaining several important parameters which describe the properties of such nuclei, notably their shape. Additional exciting subjects in current nuclear physics research are the search for the nuclear limits of stability [4] and new reaction channels as well.

## 2. Measurements with neutron generator

The first series of terbium measurements were carried out at the Nuclear Physics Department of the University of Kyiv. The determination of (*n, p*), (*n, α*), (*n, n'α*), and (*n, 2n*) cross sections was performed with the neutron-activation technique and reported earlier [5]. The cross sections were measured for (14.7 ± 0.2) MeV neutron energy. In this research three terbium disc specimens of natural composition from different producers were used. The specimens N1 (m=16 g, thickness – 8 mm) and N2 (m=2.3 g, thickness – 2 mm) originated from Russian Federation and N3 (m=2.4 g, thickness – 2 mm) – from China [6]. The neutron generator NG-300 was used as a source of (*d, d*) and (*d, t*) neutrons [7, 8]. In the first experiment the sample N1 was irradiated with (*d, t*) neutrons. To generate neutrons with energy ~ 14 MeV a molecular component of a deuteron beam was used. The target consisted of 25 Ci of tritium absorber on a Ti layer. The maximum current of deuterium ions beam was 1.0 mA. The In and Cd covering foils were used in order to suppress the scattered low-energy neutrons influence. The average flux density of (*d, t*) neutrons was ~ $1.6 \times 10^8$ ($n$/cm$^2$·s) and depended on the specimen position under irradiation. This value was experimentally determined with the $^{27}$Al($n, \alpha$)$^{24}$Na [9] ($t_{1/2}$ = 14.96 h; $E_\gamma$ = 1369 keV; $I_\gamma$ = 100% [10]) monitor reaction. Time of irradiation varied within 5 min and up to 2 hours. Afterwards with minimum cooling time the measurements of Tb specimen instrumental spectra were performed. During the procedure of one hour instrumental spectrum acquisition the 511 keV γ-line peak with essentially higher and varying intensity was observed, which was clearly expressed instead of other peaks (Fig. 1). This peak count rate equals 46 (s$^{-1}$) when the intensity of measured background's 511 keV is 0.09 (s$^{-1}$) only. Dead time was below 1%.

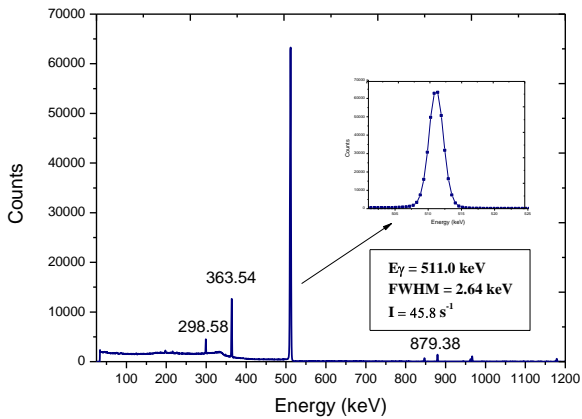

FIG. 1. Instrumental spectrum of Tb after (d-t) neutron irradiation of N1 specimen.

The FWHM of this peak was 2.64 keV (HPGe spectrometer, energy resolution - 2.0 keV for γ-ray 1,332.5 keV) which was broader than adjacent peaks and this gives us ground to indicate the case of photons detecting due to positron annihilation. For Tb specimen N2 the irradiation and measuring conditions were the same. The situation was similar to previous one, the count rate equals 35 (s$^{-1}$) and FWHM is 2.6 keV. Tb is monoisotopic abundance element and when considering the opened reaction channels at the neutron energy 14.7 MeV we could identify the following reactions: $^{159}$Tb($n, p$)$^{159}$Gd, $^{159}$Tb($n, \alpha$)$^{156}$Eu, $^{159}$Tb($n, n'\alpha$)$^{155}$Eu, $^{159}$Tb($n, 2n$)$^{158}$Tb and $^{159}$Tb($n, \gamma$)$^{160}$Tb. All products of these reactions are unstable $\beta^-$ decaying nuclei except $^{158}$Tb nucleus with $T_{1/2}$ = 180 y [10] which decays via EC and $\beta^+$ channels. To verify the energy threshold existence of the process detected, the Tb N1 specimen irradiation with (*d-d*) neutrons was performed with the neutron flux density ~ $2.6 \times 10^6$ ($n$/cm$^2$·s) and the energy of neutrons (2.8 ± 0.1) MeV. Irradiation was undertaken with the same geometry conditions and time settings as in the case of (*d-t*) neutrons. Thorough analysis of spectrum did not show any exceptional observations in the energy range around 511 keV. Due to unintelligible observation of intensity enhancing, a decision about next more deep study was undertaken. For that reason the half-life determination was done for 511 keV γ-line peak in order to find the candidate of required nucleus (source). For this aim the standard specific-activity method of examining the activity decreasing with a time was implemented. Obtained result shows a detecting at least two components with different half-life periods: $T_{1/2(1)}$ = (45 ± 10) minutes and $T_{1/2(2)}$ = (97 ± 19) minutes. The list of existing $\beta^+$ nuclei which have similar half-life values (10-60 min) and the available 511 keV γ-transitions in decay schemes was analyzed. Thus, the nucleus of $^{106}$Ag has $T_{1/2}$ = 23.3 min [10] with very intensive 511 keV γ-transition (17.03 abs.u). In order to exclude the contribution of $^{107}$Ag($n, 2n$)$^{106m}$Ag reaction the additional experiment was performed with silver specimen irradiation. The decay curves crossing of terbium and silver within first 1,000.0 s was observed and could be considered as a reliable confirmation that Ag is not a proper contributor to Tb specimen 511 keV γ-peak intensity observed.

### 3. Measurements with M-30 microtrone

Several experiments for Tb research have been performed at the Institute of Electron Physics of NAS of Ukraine based on irradiation with γ-rays in order to examine availability of mentioned effect with other type of particles (beams). The M-30 microtrone was used as a source of electrons [11]. The electron beam extracted from the accelerator was converted into bremsstrahlung with a 0.5-mm thick tantalum disk. A secondary-emission monitor was used to control the electron beam. The mean current of beam electrons was 5 μA. Energy of the accelerated electrons was controlled by the nuclear magnetic resonance method. The bremsstrahlung spectra were simulated with mcnp4c code [12] for the experiment's conditions optimizing. The average density of γ-ray flux was ~ $2 \cdot 10^9$ (1/s·cm$^2$) and this value was determined with $^{55}$Mn(γ, n)$^{54}$Mn ($t_{1/2}$ = 312.12 d; $E_\gamma$ = 834.85 keV; $I_\gamma$ = 99.97% [10]) monitor reaction, the data of cross section was taken from [13]. Tb specimen N2 was irradiated during 1.5 hour in sandwich with MnO$_2$ powder by 7.5 and then 13.5 MeV end point energy bremsstrahlung flux. This time a count rate in 511 keV peak was 23.6 (s$^{-1}$) and FWHM - 3.09 keV (Ge(Li)spectrometer, energy resolution - 3.0 keV for γ1,332. keV) for second irradiation and nothing – for first one. The occurrence of similar increased and varying 511 keV peak count rate declares the energy threshold dependence. In accordance with such assumption the sequential irradiations of Tb specimen N3 with 9.5, 11, 11.5, 12, 12.5, 16.5, and 18.5 MeV γ-rays were performed. The scheme of irradiations and measurements was similar to previous ones. Intensity increasing four times with respect to background was noticed at 12.5 MeV end point energy of bremsstranhlung spectrum. The measurements of $T_{1/2}$ were done as well and the result is presented in Fig. 2.

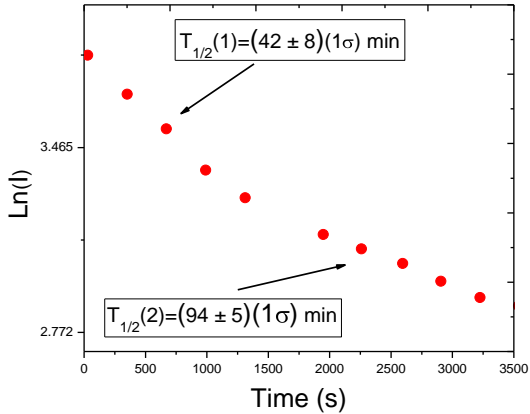

FIG. 2. Decay curve of 511keV γ-line after 12.5 MeV end point bremsstrahlung irradiation

Two component structures were observed again: similar half-life values (within uncertainty) are in agreement with corresponding ones from above. The same behavior of 511 keV peak was recorded for bremsstrahlung with 16.5 and 18.5 MeV end point energies.

## 4. Impurities and measured data analysis

Prior to irradiations all samples were checked compulsory for the presence of impurities (with radio frequency ablation and mass spectrometry analysis) [14]. Because we concluded that the products of all observed reactions could not be a suitable source of positrons the most attention was paid to impurities analysis. In Tables I-II the impurities of used Tb specimens are presented, for specimen N3 this information was provided by producer. The impurities in N2 specimen were similar to the specimen N1. For all stable isotopes of impurities all the reactions were analyzed for the subject of their reaction products, $β^+$ decay scheme and value of $T_{1/2}$ [15] accordingly to measured ones.

TABLE I. Impurities in specimen N1

| Rare-earth group | Quantity (ppm) | Element | Quantity (ppm) |
|---|---|---|---|
| La | 22.51 | Ti | 3.27 |
| Ce | 12.16 | Cr | 1.91 |
| Pr | 28.55 | Mn | 16.63 |
| Nd | 19.26 | Fe | 15.25 |
| Sm | 35.44 | Ni | 4.55 |
| Eu | – | Cu | 37.59 |
| Gd | 46.53 | Y | 270.33 |
| Dy | 706.81 | Zr | <16.74 |
| Ho | 32.31 | Nb | 0.11 |
| Er | 546.07 | Mo | 1.13 |
| Tm | 18.42 | Ag | 1.17 |
| Yb | 10.86 | Cd | 2.09 |
| Lu | 201.64 | W | 0.82 |
| *Others* | <0.01 | Pb | 16.38 |

TABLE II. Impurities in specimen N3

| Rare-earth group | Quantity (ppm) | Element | Quantity (ppm) |
|---|---|---|---|
| La | 0.1 | Sc | 0.1 |
| Ce | 0.1 | Y | 3.0 |
| Pr | 0.1 | Fe | 73 |
| Nd | 0.5 | Ca | 7 |
| Sm | 0.1 | W | 1 |
| Eu | 0.1 | Mg | 1 |
| Gd | 13.0 | Al | 3 |
| Dy | 47.0 | Ta | 10 |
| Ho | 1.0 | Si | 12 |
| Er | 1.0 | O | 200 |
| Tm, Lu | 1.0 | C | 80 |
| Yb | 0.5 | | |

It was concluded that the impurities do not cause the observation being discussed. In addition Tb specimen N3 was irradiated with (d,t) neutrons (very similar behavior of 511 keV γ peak was clearly confirmed) and the calculations of the reaction yields (upper estimates) for all impurities were done with the same final conclusions. Earlier Tb investigations were targeted on the measurements of $^{159}$Tb(n, 2n)$^{158}$Tb cross section, but the used scheme of these experiments [16, 17] did not give a possibility to observe 40 min activity of 511 keV and the authors could not see anything currently discussed. The assumption about cross section value was done and the cross section of the possible process calculated in accordance with hypotheses about some virtual nucleus formation with $T_{1/2}$ equals to the first measured component. This value was determined in reference to the reaction $^{159}$Tb$(n, p)^{159}$Gd cross section using the following expression:

$$\sigma_x = \frac{[S_0 \cdot \varepsilon_0 \cdot I_{\gamma 0} \cdot N_0 \cdot M_0 \cdot D_0 \cdot \lambda_x \cdot F_x]}{[S_x \cdot \varepsilon_x \cdot I_{\gamma x} \cdot N_x \cdot M_x \cdot D_x \cdot \lambda_0 \cdot F_0]} \cdot \sigma_0, \qquad (1)$$

where: $S$ – peak area, $\varepsilon$ – detector efficiency, $I_\gamma$ – quantum yield ($I_{\gamma x} = 2$), $N$ – isotope abundance, $M$ – mass of Tb specimen, $D$ – factor of time, $\lambda$ – decay constant, $F$ – total correction factor, $\sigma_0$ – reference cross section (subscript "0" corresponds to monitor and "x" to studied reaction accordingly). An estimated lower value of cross section for this process is (24 ± 5) mb. Since the unexplained observation has energy threshold (12.2 ± 0.3) MeV the separation energy of nucleon was analyzed for two systems with neutron and photon in entrance channel. For separation energy calculations the on-line database [18] was used, but suitable nucleus which could be formed in the $(n, 2n)$, $(\gamma, n)$ and other reactions with threshold about 12.2 MeV was not identified. The nature of interaction of neutrons and photons with nuclei is totally different. If in the case of neutrons with ~ 14 MeV energy the direct and pre-equilibrium processes occur, for photons the situation is more complicated from view point of theoretical predictions and exited states descriptions. Additional complexity is that terbium has strongly deformed nucleus and many closely spaced levels which form a double giant electric dipole resonance for photon absorption. Terbium belongs to mass range 150 < A < 190 for which the problem with parameter of deformation exists [19]. The hydrodynamic model was the first to predict the splitting of the giant E1 resonance into two components and has been quite successful in specifying the shape of the resonance. An interesting fact is that measured (12.2 ± 0.3) MeV threshold does correspond to the energy range of the largest disagreement between experimental [20, 21] and theoretical data around first maximum of GDR (Fig. 3). In Fig. 3 for comparison the data from TENDL-2009 database [22] is plotted with experimental data from EXFOR database.

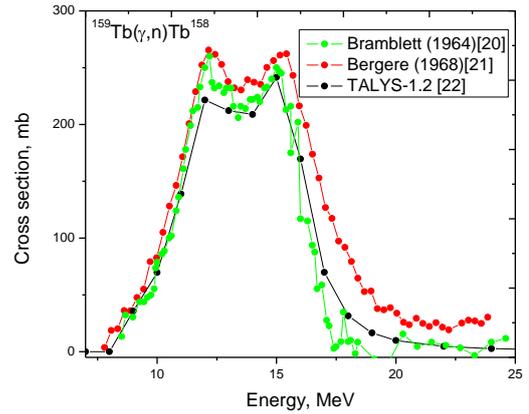

FIG. 3. The comparison of experimental and calculated data for reaction $^{159}$Tb$(\gamma,n)^{158}$Tb.

## 5. Summary

Activation measurements of terbium were performed with $(d\text{-}d)$, $(d\text{-}t)$ neutrons and bremsstrahlung spectra with end point energies ranged within (7.5-18.5) MeV. Analysis of instrumental spectra indicated unclear effect of significantly higher intensity of 511 keV γ-line what was observed at 14.7 MeV neutron energy and in (12.5-18.5) MeV bremsstrahlung end point energy range. Based on the analysis of performed experiments we do assume that the cause of wide and intensive 511 keV γ-peak formation could be related to positron emission. Such an assumption is justified by the fact that Tb is strongly deformed nucleus and due to lack of explicit theory of deformed nuclei the possible process cannot be explained yet. The essential difference between the neutron and γ-ray interaction nature was mentioned but, in fact, in both these cases the formation of the same state or may be even nucleus is occurring obviously. In accordance with recent investigation in the field of proton radioactivity and taking into account the Tb nucleus deformation even fission process at low neutron energies is possible. For finding the reliable explanation of this effect the new measurements with mono energetic neutrons are required in order to verify the threshold of this effect in the case of neutron irradiation. We can conclude that this direction of research could be important for getting a new knowledge about structure and specific properties of Tb, other rare-earth nuclei, and exotic decays with special attention to particle radioactivity in the light rare-earth region.

## 6. Acknowledgements

The authors are gratefully acknowledging all of the staff of NG-300 neutron generator group and M-30 microtrone crew for their stable and reliable operation during the experiments.